\documentstyle[12pt,epsfig]{article}
\begin{document}
\vspace{1.cm}
\begin{center}
\    \par
\    \par
\    \par
\     \par

        {\Large { \bf{  Study of Multiparticle Azimuthal Correlations in
    Central CNe, MgMg, CCu and OPb Interactions at Energy
                of 3.7 GeV per Nucleon }}}
\end{center}
\   \par
\   \par
\   \par
\   \par
\    \par
\par
{\large { \bf{L.Chkhaidze, T.Djobava, L.Kharkhelauri}}}\par
\   \par
\   \par
\    \par
{\it {High Energy Physics Institute of Tbilisi State University,
Georgia}}\par
{\it {Corresponding author's E-mail: ida@sun20.hepi.edu.ge ~~~or}}\par
{\it {\hspace{5.7cm}                 lali@sun20.hepi.edu.ge}} \par
\pagebreak
\begin{center}

                     \bf{ ABSTRACT }
\end{center}
\ \par
\par
Azimuthal correlations between protons and
between pions have been investigated in central CNe, MgMg, CCu and OPb
collisions at energy of 3.7 GeV/nucleon. "Back-to-back" (negative)
correlations have been observed for protons in CNe, CCu and for $\pi^{-}$-
mesons in  CNe and MgMg collisions. For $\pi^{-}$-mesons "side-by-side"
(positive) azimuthal correlations have been observed for heavy
systems of
CCu and OPb. The Quark Gluon String Model satisfactorily describes the
experimental results both for protons and $\pi^{-}$-mesons.
\pagebreak
\   \par
\  \par
\begin{center}
\bf { 1. INTRODUCTION }
\end{center}
\  \par
\par
Relativistic nucleus--nucleus collisions are very well suited for
investigation of excited nuclear matter properties which are the
subject of intense studies both experimentally and theoretically.
Theoretical models predict formation of exotic states of nuclear
matter, for example the phase transition to a quark-gluon plasma [1,2].
One of the main goals of relativistic heavy-ion collision
experiments is to study nuclear matter under extreme conditions of high
densities and temperatures. The most impressive results of high energy
heavy ion research so far  are new collective phenomena discovered in
these reactions. Study of multiparticle correlations offers unique
information about space-time evolution of the collective system. During
last
years
an intensive analysis of experimental data have been carried out using the
collective variables, which depend on the momentum of all secondary particles,
to reveal a nontrivial effects in nucleus--nucleus collisions.
nucleus--nucleus collisions.
\par
Experimental discovery of such transitions is impossible without
understanding of the mechanism of collisions and studying the
characteristics
of multiparticle production in nucleus-nucleus interactions. Multiparticle
correlations had been
investigated at the first time at the BEVALAC more than 10 years ago
[3].
\par
In this article we present results of the analysis of multiparticle
correlations in central CNe, MgMg, CCu and OPb collisions at energy
of 3.7 GeV/nucleon. Azimuthal correlations between protons and between
pions and dependence of these correlations on the projectile ($A_{P}$)
and target ($A_{T}$) nucleus have been investigated.
\  \par
\  \par
\begin{center}
\bf { 2. EXPERIMENT }
\end{center}
\  \par
\par
Data were obtained using the 4$\pi$  spectrometer
SKM-200 -- GIBS of the Dubna JINR [4,5]. The SKM-200 -- GIBS set-up
 consists of a 2 m
streamer chamber with fiducial volume of a 2$\times$1$\times$0.6 m$^{3}$,
placed
in a magnetic
field of $\sim$ 0.8 T ($\sim$ 0.9 T for MgMg) and a triggering system.
The streamer chamber was exposed to beams of C, O and Mg nuclei accelerated in the
synchrophasotron up to energy of 3.7 GeV per incident nucleon.
Solid targets in the form of thin discs with thickness (0.2$\div$0.5)
g/cm$^{2}$
(the thickness of Mg was 1.5 g/cm$^{2}$; neon-gas filling of the
chamber also served as a nuclear target) were mounted inside the chamber
at a distance of 70 cm from the entrance window and at a height
of 8 cm above the middle electrode. Photographs of the events were
taken using an optical system with 3 objectives. Experimental set-up
and the logic of the triggering system are presented in Fig. 1.
Triggering system allowed the selection of "inelastic" and "central"
collisions.
\par
   The "inelastic" trigger, consisting of two sets of scintillation counters
mounted upstream ($S_{1}$ -- $S_{4}$) and downstream ($S_{5}$,
$S_{6}$) the
chamber, has been selected all inelastic interactions of incident nuclei
on a target.
\par
   The "central" triggering system was consisting of the same upstream
part as
in the "inelastic" system and of scintillation veto counters
($S_{\rm{ch}}$, $S_{\rm{n}}$), to reject a projectile and its
charged and neutral spectator fragments, in the downstream part. All
counters were made from plastic scintillators and worked with
photomultipliers PM-30. The $S_{1}$ counter with the scintillator of
20$\times$20$\times$0.5 cm$^{3}$ size, worked in the amplitude regime
and identified the beam nuclei by its charge.  The nuclei from the beam,
going to the target, have been selected using the profile counters
$S_{2}$, $S_{3}$ with the plastic of 0.15 mm diameter and 3 mm thickness
and "thin" counter $S_{4}$ (15 mm and 0.1 mm correspondingly). The
$S_{\rm{ch}}$ counters (two counters with plastic of
40$\times$40$\times$0.5 cm$^{3}$ size) were placed at a distance of 4 m
downstream from the target and registered secondary charged particles,
emitted from the target within a cone of half angle $\theta_{\rm{ch}}$
=~2.4~grad or 2.9~grad. The $S_{\rm{n}}$ counters were registering the
neutrons, emitted from the target in the same solid angle
$\theta_{\rm{n}}$ =~2.4~grad or 2.9~grad..  The $S_{\rm{n}}$ telescope
consisted of the five counters of 40$\times$40$\times$2 cm$^{3}$ size,
layered by 10 cm thick iron blocks.  The central trigger was selecting
events defined as those with no charged projectile
spectator fragments and spectator neutrons ($p$/$Z$ $>$ 3 GeV/$c$)
emitted at angles $\theta_{\rm{ch}}$ = $\theta_{\rm{n}}$ =~2.4 grad or
2.9~grad.  ($\sim$ 4msr). The  trigger efficiency was 99~$\%$ and
80~$\%$ for charged and neutral projectile fragments, respectively.  The
trigger mode for each exposure is defined as Tr($\theta_{\rm{ch}}$,
$\theta_{\rm{n}}$) ($\theta_{\rm{ch}}$ and $\theta_{\rm{n}}$ expressed
in degrees and rounded to the closest integer value).  Thus
nucleus--nucleus interactions obtained with this set-up correspond to
the following Tr($\theta_{\rm{ch}}$, $\theta_{\rm{n}}$) triggers:  CNe
-- Tr(2, 0), MgMg -- Tr(2, 2), CCu -- Tr(2, 0) and Tr(3, 3), OPb --
Tr(2, 0).  \par Biases and correction procedures were discussed in
detail in [4,5].  The ratio  $\sigma_{\rm{cent}}$/$\sigma_{\rm{inel}}$
(that characterizes the centrality of selected events) is (9$\pm$1)~$\%$
for CNe and (21$\pm$3)~$\%$ for CCu and the fraction of central MgMg
events is $\approx$ 4$\times$10$^{-4}$ among all inelastic interactions.
Average momenta measurement errors
$<\Delta p/p>$ = (8$\div$10)~$\%$ for protons and 5~$\%$ for pions,
corresponding errors of the production angles are
$\Delta$$\theta$ = (1$\div$2)~grad. and ~0.5~grad.
( for $\pi^{-}$-mesons in MgMg interactions we had
$<\Delta p/p>$ =~1.5~$\%$, $\Delta$$\theta$ =~0.3~grad.)
\  \par
\  \par
\begin{center}
\bf { 3.  QUARK GLUON STRING MODEL }
\end{center}
\  \par
\par
Several theroretical models of nucleus--nucleus collisions at high
energy have been proposed in Refs. [6-8]. These models allow one to test
various assumptions concerning the mechanism of particle production at
extreme conditions achieved only in nucleus--nucleus collisions.
The Quark Gluon String Model (QGSM) has been used for the comparison
with our experimental results. QGSM is presented in detail in papers
[9,10]. The QGSM is based on the Regge and string phenomenology of
particle
production in inelastic binary hadron collisions. For describtion the
evolution
of the hadron and quark-gluon phases, uses a coupled system of
Boltzmann-like kinetic equations. Nuclear collisions are treated
as a mixture of independent interactions of the projectile and target
nucleons, stable hadrons and short lived resonances. The QGSM includes
low mass vector mesons and baryons with spin 3/2, mostly
$\Delta$(3/2, 3/2) via resonant reactions. Pion absorption by $NN$
quasideuteron pairs is also taken into account. The coordinates of
nucleons are generated according to a realistic nuclear density.
The sphere of the nucleus is filled with the nucleons at a
condition that the distance between them is greater than 0.8 fm.
The nucleon momenta are distributed in the range of
0 $\leq$ $p$ $\leq$ $p_{\rm{F}}$. The maximum nucleon Fermi momentum is
\par
\begin{center}
   $p_{\rm{F}}$ = $(3\pi^{2})^{1/3}h\rho^{1/3}(r)$ \hspace{4cm}(1)
\end{center}
 where $h$=0.197 fm$\cdot$GeV/$c$, $\rho(r)$ is nuclear density.
\   \par
\par
The procedure of event generation consists of three steps: the definition of
configurations of colliding nucleons, production of quark-gluon strings
and fragmentation of strings (breakup) into observed hadrons. The model
includes also the formation time of hadrons. The QGSM has been extrapolated
to the range of intermediate energy (${\sqrt{s}}$ $\leq$ 4 GeV)
to use it as a basic process during the generation of
hadron-hadron collisions. Masses of "strings"
produced at ${\sqrt{s}}$ = 3.6 GeV were small (usually not greater
then 2 GeV), and they were fragmenting mainly ($\approx$~90~$\%$)
through two-particle decays. For main $NN$ and $\pi$$N$ interactions the
following topological quark diagrams [7] were used: binary, "undeveloped"
cylindrical, diffractive and planar. The binary process makes a main
contribution which is proportional to 1/$p_{\rm{lab}}$. It corresponds
to quark
rearrangement without direct particle emission in the string decay. This
reaction predominantly results in the production of resonances (for
instance, $ p + p$ $\rightarrow$ $N + $ $\Delta^{++}$), which are
main source of pions. The comparable contributions to the
inelastic cross section, which however
decreases with decreasing $p_{\rm{lab}}$, come from the diagrams
corresponding
to the "undeveloped" cylindrical diagrams and from the diffractive processes.
Transverse momenta of pions produced in quark-gluon string fragmentation
processes are the product of two factors: string motion on the whole as a
result of transverse motion of constituent quarks and
$q$$\bar{q}$ production in string breakup. The transverse
motion of quarks inside hadrons was described by the Gaussian
distribution with variance $\sigma^{2}$~$\approx$~0.3~(GeV/$c$)$^{2}$.
Transverse  momenta k$_{T}$ of produced $q$$\bar{q}$
in the c.m.s. of the string follow the dependence:
\  \par
\begin{center}
$W(k_{T}) = 3B/\pi(1+Bk^{2}_{T})^{4}$    \hspace{3cm}    (2)
\end{center}
where $B$ = 0.34 (GeV/$c$)$^{-2}$.
\par
 The cross section of hadron interactions were taken from experiments.
Isotopic invariance and predictions of the additive quark model [11]
(for meson--meson cross sections, etc.) were used to avoid data
deficiency. The resonance cross sections were assumed to be
identical to the stable particle cross sections with the same
quark content. For the resonances the tabulated  widths were used.
\par
The QGSM simplifies the nuclear effects. In particular, coupling of nucleons
inside the nucleus is neglected, and the decay of excited recoil nuclear
fragments and coalescence of nucleons is not included.
\par
We have generated CNe, CCu, OPb and MgMg interactions using the Monte
Carlo generator COLLI, which is based on the QGSM and then traced through the
detector and trigger filter.
In the generator COLLI there are two possibilities to generate events:
1) at not fixed impact parameter $\tilde{b}$ and  2) at fixed $b$.
From the impact parameter distributions we obtained the mean value
of $<b>$ = 2.2 fm, 1.3 fm, 2.7 fm and 3.7 fm
for CNe, MgMg, CCu and OPb collisions. For the obtained
values of $<b>$ we have generated a total sample of events
6270, 9320, 2430 and 6200, respectively.
\par
The QGSM overestimate the production of low momentum protons with
$p<$ 0.2 GeV/$c$, which are mainly the target fragments and were
excluded
from
the  analysis. From the analysis of generated events protons with deep
angles greater than 60~grad. had been excluded, because in the experiment
the registration efficiency of such vertical tracks is low.
\  \par
\  \par
\begin{center}
\bf { 4.  AZIMUTHAL CORRELATIONS BETWEEN PROTONS AND BETWEEN PIONS }
\end{center}
\  \par
\par
In Refs. [12,13] the procedure to study of the correlation between groups
of particles has been developed. The azimuthal
correlation function was defined by the relative opening angle
between the transverse momentum vector sums
of particles emitted forward and backward with respect to the rest
frame of the target nucleus ($y_{t}$ =~0.2).
\par
We have applied this method for our data, but the analysis have been
carried
out in the central rapidity region instead of the target rapidity range
of Refs. [12,13]. The analysis have been performed event by event, in
each event we denote the vectors:
\  \par
\begin{center}
${\bf{Q}}_{B}=\sum\limits_{y_{i}<<y>}{\bf{P}}_{{\perp}i}$
\hspace{3cm}    (3)
\end{center}
and
\  \par
\begin{center}
${\bf{Q}}_{F}\sum\limits_{y_{i}\geq<y>}{\bf{P}}_{{\perp}i}$
\hspace{3.3cm}    (4)
\end{center}
where $<y>$ is the average rapidity in each event.
\par
Then the correlation function C($\Delta \varphi$) is constructed as follows:
\  \par
\begin{center}
C($\Delta \varphi$) = dN/d$\Delta \varphi$  \hspace{6cm}    (5)
\end{center}
where $\Delta \varphi$ is the angle between the
vectors ${\bf{Q}}_{B}$ and ${\bf{Q}}_{F}$:
\  \par
\begin{center}
$\Delta
\varphi$ = arccos(${\bf{Q}}_{B}\cdot{\bf{Q}}_{F}) /
\vert{\bf{Q}}_{B}\vert\cdot\vert{\bf{Q}}_{F}\vert$.
\hspace{3cm}    (6)
\end{center}
\par
Essentially, C($\Delta \varphi$) measures whether the particles in the
backward and forward hemispheres are preferentially emmited "back-to back"
($\Delta \varphi$~=~180~grad.) or "side-by-side" ($\Delta \varphi$=
~0~grad.) [12]. The protons from CNe and CCu collisions have been analysed
with use of this method.
\par
For the analysis it is necessary to perform an identification of
$\pi^{+}$ mesons, the admixture of which
amongst the charged positive particles is about (25$\div$27)$\%$ .
The statistical method have been used for identification of $\pi^{+}$ mesons.
The main assumption is based on the similarity of spectra of
$\pi^{-}$ and $\pi^{+}$ mesons ($n_{\pi}$, $p_{T}$,
$p_{L}$). The two-dimentional
distribution of ($p_{T}$, $p_{L}$) variables have been used
for identification
of $\pi^{+}$ mesons. The whole plane is divided into 7 zones. For example,
for CNe collisions:\\
1) $p_{L}$ $>$ 2.5  GeV/$c$ or $p_{T}$ $>$ 0.9  GeV/$c$; \\
2) 0 $\leq$ $p_{L}$ $\leq$ 1.4 GeV/$c$ and $p_{T}$ $\leq$ 0.7
 GeV/$c$ -- PMAX;\\
3) 0 $\leq$ $p_{L}$ $\leq$ 1.4 GeV/$c$ and $p_{T}$ $>$ 0.7
 GeV/$c$;\\
4) 1.4  GeV/$c$ $<$ $p_{L}$ $<$  2.5  GeV/$c$;\\
5) -0.2 $\leq$ $p_{L}$ $\leq$ 0 GeV/$c$ and $p_{T}$ $\leq$
0.3 GeV/$c$;\\
6) -0.2 $\leq$ $p_{L}$ $\leq$ 0 GeV/$c$ and
$p_{T}$ $>$ 0.3 GeV/$c$;\\
7) $p_{L}$ $<$ -0.2  GeV/$c$.\\
The zone 2 of maximal overlap -- PMAX, in its turn, is divided into
7$\times$7=49 cells. The probability of hitting
each zone and respectively the cell
by $\pi^{-}$ mesons and charged positive particles  is
defined and the relative probability of hitting of $\pi^{-}$ mesons is calculated
as a result. The admixture of $\pi^{+}$ mesons in the zone 1 is negligible.
 The procedure of dividing the ($p_{T}$, $p_{L}$) plane into
   the cells allows to simplify the mathematical algorithm and improves the
accuracy of the identification. In Fig. 2 the division of the
($p_{T}$, $p_{L}$) plane is presented for CNe collisions. It
was
assumed, that $\pi^{+}$  and $\pi^{-}$ mesons hit given cell with equal
probability ( the equal probability densities for $\pi^{+}$ and $\pi^{-}$ were
assumed).
\par
The identification in fact is equal to summing of  hitting
probabilities into each cell and when the sum reaches the critical
value, the particle is considered as $\pi^{+}$ meson. The rest
of particles are assumed to be protons. For each proton and $\pi^{+}$
meson the sign is recorded on
DST (Data Summar Tape), which indicates to which zone of ($p_{T}$,
$p_{L}$)
plane belongs
the given particle and what is the  probability. Particles
with $p_{T}$ $>$ 0.9 GeV/$c$ or $p_{L}$ $>$ 2.5 GeV/$c$ are
unambiguous protons with probability equal to 1.
\par
After identification of $\pi^{+}$ mesons in the event, the difference of
$\pi^{+}$ and $\pi^{-}$ mesons $\Delta n$ is determined. If
$|\Delta n| > 2$, in the region PMAX the identification
$\pi^{+}$ $\leftrightarrow$ proton is interchanged and for that reason
particles with smaller probability are chosen. If the condition of approachement
of multiplicities is not fulfiled, then  in this case
into the "head" information of the event such a sign is recorded,
 which allows to exclude the given
event from the further analysis.
\par
After performed identification the admixture of $\pi^{+}$ mesons amongst the
protons is not exceeding (5 $\div$ 7)$\%$.
The mean values of multiplicity, momentum and transverse momentum
for $\pi^{+}$ and $\pi^{-}$ mesons are presented in Tabl. 1.
One can see, that the average kinematical
characteristics of $\pi^{+}$ and $\pi^{-}$ mesons
coincide within the errors satisfactorily.
\par
The numbers of events for CNe and
CCu collisions and the mean rapidities of analysed protons $<y>$
are listed in Tabl. 2.
Fig. 3 shows the experimental correlation function C($\Delta \varphi$)
for protons from central CNe and CCu collisions. One can observe from
Fig. 3 a clear correlation for protons (correlation increases with
$\Delta \varphi$, reaches maximum at $\Delta \varphi$ = 180~grad.). To
quantify these experimental results, the data were fitted by:
\  \par
\begin{center}
C($\Delta \varphi$) = 1 +$\xi cos(\Delta \varphi$).  \hspace{5.cm}    (7)
\end{center}
Results of the fitting are listed in Tabl. 2. The strength of the
correlation
is defined as
\  \par
\begin{center}
$\zeta$ = C(0~grad.)/C(180~grad.) = (1 + $\xi$)/(1 -
$\xi$).  \hspace{2.7cm}
(8)
\end{center}
As it can be seen from the Tabl. 2, the asymmetry coefficient $\xi$ $<$ 0
and thus
the strength of correlation $\zeta$ $<$ 1 for protons in both CNe and CCu
interactions, meaning the negative correlations and that protons are
preferentially emitted back-to-back. Absolute values of $\xi$
increase and $\zeta$ decrease with target mass increases.
\par
A similar negative (back-to-back) correlation have been observed by
Plastic Ball collaboration at Bevalac between the "slow" (40 $<$ E $<$ 240 MeV)
and the "fast" (E $>$ 240 MeV) fragments for symmetric
($^{40}$Ca + $^{40}$Ca, $^{93}$Nb + $^{93}$Nb) and
asymmetric ($^{20}$Ne + $^{93}$Pb) pairs of nuclei in the energy interval
of 0.4 to 1 GeV/nucleon [14,15] and also between protons in $p$ + Au
collisions
at energy of 4.9 GeV/nucleon [12,13].
The investigation of large angle two-particle correlations [16],
have been carried out at Dubna for collisions of $^{4}$He
and $^{12}$C with different nuclear
targets ($^{27}$Al, $^{64}$Cu and $^{93}$Pb) at energy of 3.6
GeV/nucleon. For protons and deuterons,
negative (back-to-back) correlation was observed for all targets. In
CC inelasic interactions at a momentum of 4.2 GeV/$c$/nucleon in the
2-meter
Propan Bubble Chamber of JINR [17] the back-to-back
azimuthal correlations
between the groups of the particles (protons) emitted in the forward and
backward hemispheres in the c.m.s. of the collisions (see
Fig. 5$a$ [17]) have been obtained.
Protons showed a typical back-to-back (negative) correlation in the
$p$ -, O -, and S - induced reactions on different nuclei
(Au, Ag, Al, C) at the CERN-SPS (WA80 collaboration) at energy
of 60 and 200 GeV/nucleon [12,13]. Azimuthal correlations in the
target
rapidity range of 0.1 $\leq$ $y_{0}$ $\leq$ 0.3 have been
obtained and within
these limits no significant change of the correlation functions has been
observed. We have studied the strength of the
correlation functions in central CNe and CCu collisions
for different rapidities ($y_{0}$) and emission
angle ($\theta$) intervals in the l. s. (see Tabl. 3).
One can see from the Tabl. 3 that absolute values of asymmetry
coefficient decrease from central rapidity region to the target
fragmentation range.
\par
The back-to back (negative) emission of protons can be understood as
resulting from
(local) total momentum conservation [13]. This behaviour is in a good
agreement with collective nuclear matter flow concept [16].
\par
In view of the strong coupling between the nucleons and pions, it is
interesting to know the origin of correlations between pions. The
$\pi^{-}$-mesons
in our experiment have been identified practically unambiguously, the
admixture of e$^{-}$, $\bar{p}$ and $K^{-}$-mesons is almost negligible
[4].
We have studied also correlations between $\pi^{-}$-mesons. For CNe
interactions into the analysis $\pi^{+}$ mesons have been included also in
order
to increase the multiplicity in each event. Correlation functions for
$\pi^{-}$-mesons in CNe, MgMg, CCu and OPb interactions are presented
in Fig. 4. One can observe from Fig. 4$a$ a clear
back-to-back ($\xi$ $<$ 0,
$\zeta$
$<$ 1, i.e. negative) correlations for pions for light system of CNe
(analogy as for the protons in CNe collisions). Study of
interactions of the symmetric pair of nuclei MgMg
(6239 collisions, 50775 $\pi^{-}$-mesons)
gives the possibility of a better manifestation of nuclear effects then
for the asymmetric pairs of nuclei. For MgMg collisions a back-to-back pion
correlations had been obtained only for the events with multiplisity
$n_{-}$ $>$ 7 and no correlations for $n_{-}$ $\leq$ 7 (Fig. 4$a$).
\par
For heavy, asymmetric pairs of nuclei CCu and OPb the side-by-side
($\xi$ $>$ 0 and $\zeta$  $>$ 1, i.e. positive) correlations of pions
can be seen from Fig. 4$b$. Similar, side-by-side correlations of
pions have
been observed in $p$~+~Au collisions at Bevalac (4.9 GeV/nucleon) and
CERN-SPS
(60 and 200 GeV/nucleon) energies [12, 13]. These results agree with
that of Refs. [18, 19]. Large angle two-particle correlations carried
out at energy of 3.6 GeV/nucleon at JINR [18] for $^{4}$He- and
$^{12}$C-beams, showed the negative (back-to-back) pion correlations for
light (Al) and the positive (side-by-side) correlations for the heavy
target (Pb) (no correlation for a medium target Cu).
\par
One can see
from Tables 2 and 4 the absolute values of the asymmetry coeficient
($\mid\xi\mid$) increase and the strength of correlations ($\zeta$)
decrease while target mass increases for both protons and pions
back-to-back (negative) correlations in contrast to the results of Ref.
[13], where back-to-back asymmetry of protons tends to vanish with
increase of target mass in proton induced reactions at 200 GeV/$c$ (see
Fig. 4 [13]). For side-by-side (positive) correlations of pions in CCu
and OPb collisions $\xi$ and  $\zeta$ increase with the target mass due
to the increasing amount of matter in their path.  \par The reason for
the observed difference behaviour between protons and pions comes from
the pion absorption in the excited target matter ($\pi$ + $N$
$\rightarrow$ $\Delta$ and $\Delta$ + $N$  $\rightarrow$ $N$ + $N$) [12,
13]. While the back-to-back emission of protons can be understood as
resulting from transverse momentum conservation, the pion correlations
show, in the data, an opposite behaviour. The side-by-side correlation
of pions can naturally be explained based on the picture that pions,
which are created in collision at a b$\neq$0 fm ( b is the impact
parameter) either suffer rescattering or even complete absorption in the
target spectator matter. Both processes will result in a relative
depletion of pions in the geometrical direction of the target spectator
matter and hence will cause an azimuthal side-by-side (positive) correlation
as observed in the experimental data. This picture is further supported by
calculations within the framework of the RQMD model [20] , which
includes pion absorption by
excited nuclear matter based on experimentally measured cross sections.
\par
The QGSM yields a significant azimuthal
correlations, which follow trends similar to the experimental data
(Figs. 3, 4).
\par
To be convinced, that the azimuthal correlations in Figs. 3, 4 (for both
experimental and QGSM data) between protons and between pions are due to
correlations between these particles and can not be the result of
detector
biases or finite-multiplicity effects, we obtained data for dependence
of C($\Delta \psi$)
on $\psi$ for secondaries, where $\psi$ is the angle between the transverse
momentum of each particle emitted in the backward (forward) hemisphere and
${\bf{Q}}_{B}$ (${\bf{Q}}_{F}$) vector,
respectively. One can see from Fig. 5, that there is no
correlation for CCu interactions both for protons (Fig. 5$a$)
and for pions (Fig. 5$b$). Similar results have been obtained for CNe
collisions too.
\par
As obtained in our previous articles [21--23], the
dependence of the mean transverse momentum in the reaction plane $<P_{X}>$
on the normalized rapidity $y/y_{p}$ in the l. s. showed the typical
$S$-shape behaviour in CNe and CCu
collisions for protons and pions. For CNe collisions $<P_{X}>$  for pions
is directed in the same direction as for protons, i.e. flows of protons and
pions are correlated, while for CCu interactions the $<P_{X}>$ of
$\pi^{-}$-mesons is directed oppositely to that of the protons (antiflow)
(see Fig. 1 [22]). In MgMg central collisions [24], for
$\pi^{-}$-mesons
with multiplicity $n_{-} >$ 7 the dependence of $<P_{X}(Y)>$
on $Y$ exhibits $S$-shape behaviour similar to the form of the $<P_{X}>$
spectra for protons and pions in central CNe collisions.
\par
In order to extend these investigations, we have obtained the relation
between $<P_{X}>^{2}$ and the angle $\varphi$, where $\varphi$ is the
opening angle
between ${\bf{Q}}_{B}$ and  ${\bf{Q}}_{F}$ vectors.
One can see from Fig.~6, that for protons in CNe and CCu collisions,
the distributions show $S$-shape behaviour and slopes of
distributions increase with target mass. One can see from Fig. 6 that
at $\varphi$ = 90~grad. the values of  $<P_{X}>$ not depend on
A$_{T}$.
\  \par
\  \par
\  \par
\  \par
\begin{center}
\bf { 4.  CONCLUSION }
\end{center}
\  \par
\par
The study of azimuthal correlations between protons and between pions
in
central CNe, MgMg, CCu and OPb collisions have been carried out.
\par
1. For protons a "back-to back" (negative) correlations were observed in
CNe and CCu interactions. The asymmetry coefficient $\xi$ ($\xi$ $<$ 0)
increases and the strength of correlation $\zeta$ ($\zeta$ $<$ 1) decreases
with increase of the target mass.
\par
2. A "back-to-back" pion correlations had been obtained for a light,
symmetric
pairs of nuclei (CNe and MgMg), where $\xi$ and $\zeta$ parameters have
the same behaviour as for protons.
\par
3. For heavy pairs of nuclei (CCu and OPb) "side-by-side" (positive)
pion correlation were observed. The asymmetry coefficient ($\xi$ $>$ 0) and
the strength of correlations ($\zeta$ $>$ 1) increase with increase of
projectile ($A_{P}$) and target ($A_{T}$) mass.
\par
4. The dependence of the square of the mean transverse momentum in the
reaction
plane $<P_{X}>^{2}$ on $\varphi$ (the angle between the vector sums of
the forward and
backward emitted particles) shows S-shaped behaviour.
Slopes of distributions increase with target mass.
\par
5. The QGSM satisfactorily describes azimuthal correlations of protons
and $\pi^{-}$-mesons for all pairs of nuclei.
\   \par
\   \par
\   \par
ACKNOWLEDGEMENTS
~~ We would like to thank M.Anikina, A.Golokhvastov, S.Khorozov and
J.Lukstins
for  fruitful collaboration during the obtaining of the data.
We are very grateful to Z. Menteshashvili for reading the manuscript.
\pagebreak
\par

\par
\newpage
\begin{center}
\bf{FIGURE CAPTIONS}
\end{center}
\   \par
{\bf{Fig.1.}}
 Experimental set-up. The trigger and trigger distances are
not to scale.
\   \par
{\bf{Fig.2}}
Two dimentional ($p_{T}$, $p_{L}$) ($p_{T}$ and $p_{L}$ denote
the transverse and longitudinal momenta, correspondently) plot
for the
identificaton of protons and $\pi^{+}$ mesons in CNe collisions.
\   \par
{\bf{Fig.3.}}
The dependence of the correlation function C($\Delta \varphi$) on the
$\varphi$ for protons from: ($\circ$) -- CNe and ($\bigtriangleup$) --
CCu collisions. ({\scriptsize$\times$}) -- the QGSM data. Dashed curves are the
results of the approximation of the data (see text).
\   \par
{\bf{Fig.4.}}
The dependence of the correlation function C($\Delta \varphi$) on the $\varphi$
for $\pi^{-}$-mesons from: (a) ($\circ$) -- CNe, ($\bullet$) -- MgMg
($n_{-}$ $>$ 7),
  ($\diamond$) -- MgMg ($n_{-}$ $\leq$ 7) and (b) ($\bigtriangleup$) --
C-Cu,
({\scriptsize$\oplus$}) -- OPb collisions.
({\scriptsize$\times$}, $\star$) -- the QGSM data, correspondently.
Solid and dashed curves are the results of the approximation of the
data
(see text).
\   \par
{\bf{Fig.5.}}
The dependence of the correlation function C($\Delta \psi$) on the $\psi$
 in CCu collisions for protons (a) and for $\pi^{-}$-mesons (b).
($\circ$) -- for particles emitted in the backward hemisphere,
($\bigtriangleup$) -- for particles emitted in the forward hemisphere.
({\scriptsize$\times$}) -- the QGSM data.
Curves are the results of the approximation of the data by the first
order polynoms.
\   \par
{\bf{Fig.6.}}
The dependence of $< P_{x} >^{2}$ on $\varphi$ (as described in the text)
for protons in: ($\circ$) -- CNe and ($\bigtriangleup$) -- CCu
collisions.
({\scriptsize$\times$}, $\star$) -- the QGSM data, correspondently.
Solid curves are the results of the approximation of the data by 4th
order polynoms.
\newpage
Table 1. The number $\pi^{+}$ and $\pi^{-}$ mesons and their
average kinamatical characteristics in CNe and CCu collisions
after the identification of protons
and $\pi^{+}$ mesons.
\   \par
\   \par
\   \par
\   \par
\   \par
\   \par
\begin{tabular}{|l|c|c|c|c|c|}    \hline
&&&&&   \\
$A_{P} - A_{T}$& Particle& $N_{\pi}$&$<n_{\pi}>$& $<p_{T}>$ & $<p>$\\
	       & Type    &          &          & GeV/$c$     & GeV/$c$\\
&&&&&   \\
\hline
&&&&&   \\
  & $\pi^{+}$ & 3089     & 4.26 $\pm$ 0.06 &0.234 $\pm$ 0.003 &
0.600 $\pm$ 0.009     \\
&&&&&   \\
\cline{2-6}
 CNe&&&&&   \\
 & $\pi^{-}$   & 3120    & 4.31 $\pm$ 0.07 &0.226 $\pm$ 0.005 &
0.612 $\pm$ 0.009     \\
&&&&&   \\
\hline
&&&&&   \\
&  $\pi^{+}$ & 3713 & 5.74 $\pm$ 0.09 &0.222 $\pm$ 0.003 &
0.522 $\pm$ 0.008     \\
&&&&&   \\
\cline{2-6}
 CCu&&&&&   \\
 & $\pi^{-}$   & 3635    & 5.68 $\pm$ 0.10 &0.213 $\pm$ 0.005 &
0.508 $\pm$ 0.009     \\
&&&&&   \\
\hline
\end{tabular}
\newpage
Table 2. The number of experimental events ($N_{\rm{event}}$) and
participant protons ($N_{\rm{prot.}}$), the asymmetry coefficient
($\xi$), the strength of the correlation ($\zeta$) and the mean
rapidity ($<y>$) of protons.\\
\   \par
\   \par
\   \par
\   \par
\   \par
\   \par
\begin{tabular}{|c|c|c|c|c|c|}    \hline
&    &     &    &     &   \\
$A_{P} - A_{T}$ &$N_{\rm{event}}$ & $N_{\rm{prot.}}$ & $\xi$ & $\zeta$ &
$<y>$ \\
\hline
&    &     &    &     &   \\
CNe  & 723 &  9201  & -0.23 $\pm$ 0.05 & 0.63 $\pm$ 0.09 & 1.07
$\pm$ 0.07\\
\hline
&    &     &    &     &   \\
CCu  & 663 & 12715  & -0.35 $\pm$ 0.05 & 0.48 $\pm$ 0.06 & 0.73
$\pm$ 0.05\\
\hline
\end{tabular}
\newpage
Table 3. Values of the coefficient of asymmetry ($\xi$) for protons
in CNe and CCu collisions for different intervals of emmision angle
($\theta$) and rapidity ($y_{0}$).
\   \par
\   \par
\   \par
\   \par
\   \par
\   \par
\begin{tabular}{|c|c|c|c|c|}     \hline
\multicolumn{1}{|c|}{coeff.}&
\multicolumn{2}{|c|}{}&
\multicolumn{1}{|c|}{}&
\multicolumn{1}{|c|}{}\\
\multicolumn{1}{|c|}{of asymm.}&
\multicolumn{2}{|c|}{range}&
\multicolumn{1}{|c|}{CNe}&
\multicolumn{1}{|c|}{CCu}\\
\hline
\multicolumn{1}{|c|}{}&
\multicolumn{2}{|c|}{in central region}&
\multicolumn{1}{|c|}{}&
\multicolumn{1}{|c|}{}\\
\multicolumn{1}{|c|}{}&
\multicolumn{2}{|c|}{0.5 $\leq$ $y_{0}$ $\leq$ 2.5}&
\multicolumn{1}{|c|}{-0.23 $\pm$ 0.05}&
\multicolumn{1}{|c|}{-0.35 $\pm$ 0.05}\\
\cline{2-5}
\multicolumn{1}{|c|}{}&
\multicolumn{1}{|c|}{}&
\multicolumn{1}{|c|}{}&
\multicolumn{1}{|c|}{}&
\multicolumn{1}{|c|}{}\\
\multicolumn{1}{|c|}{}&
\multicolumn{1}{|c|}{}&
\multicolumn{1}{|c|}{$y_{0}$=0.2}&
\multicolumn{1}{|c|}{--}&
\multicolumn{1}{|c|}{-0.22 $\pm$ 0.05}\\
\cline{3-5}
\multicolumn{1}{|c|}{}&
\multicolumn{1}{|c|}{10~grad $\leq$ $\theta$ $\leq$ 180~grad.}&
\multicolumn{1}{|c|}{}&
\multicolumn{1}{|c|}{}&
\multicolumn{1}{|c|}{}\\
\multicolumn{1}{|c|}{}&
\multicolumn{1}{|c|}{}&
\multicolumn{1}{|c|}{$y_{0}$=0.3}&
\multicolumn{1}{|c|}{-0.11 $\pm$ 0.03}&
\multicolumn{1}{|c|}{-0.27 $\pm$ 0.05}\\
\cline{2-5}
\multicolumn{1}{|c|}{$\xi$}&
\multicolumn{1}{|c|}{}&
\multicolumn{1}{|c|}{}&
\multicolumn{1}{|c|}{}&
\multicolumn{1}{|c|}{}\\
\multicolumn{1}{|c|}{}&
\multicolumn{1}{|c|}{}&
\multicolumn{1}{|c|}{$y_{0}$=0.2}&
\multicolumn{1}{|c|}{--}&
\multicolumn{1}{|c|}{-0.13 $\pm$ 0.05}\\
\cline{3-5}
\multicolumn{1}{|c|}{}&
\multicolumn{1}{|c|}{20~grad $\leq$ $\theta$ $\leq$ 180~grad.}&
\multicolumn{1}{|c|}{}&
\multicolumn{1}{|c|}{}&
\multicolumn{1}{|c|}{}\\
\multicolumn{1}{|c|}{}&
\multicolumn{1}{|c|}{}&
\multicolumn{1}{|c|}{$y_{0}$=0.3}&
\multicolumn{1}{|c|}{--}&
\multicolumn{1}{|c|}{-0.19 $\pm$ 0.05}\\
\cline{2-5}
\multicolumn{1}{|c|}{}&
\multicolumn{1}{|c|}{}&
\multicolumn{1}{|c|}{}&
\multicolumn{1}{|c|}{}&
\multicolumn{1}{|c|}{}\\
\multicolumn{1}{|c|}{}&
\multicolumn{1}{|c|}{}&
\multicolumn{1}{|c|}{$y_{0}$=0.2}&
\multicolumn{1}{|c|}{--}&
\multicolumn{1}{|c|}{-0.07 $\pm$ 0.01}\\
\cline{3-5}
\multicolumn{1}{|c|}{}&
\multicolumn{1}{|c|}{30~grad $\leq$ $\theta$ $\leq$ 180~grad.}&
\multicolumn{1}{|c|}{}&
\multicolumn{1}{|c|}{}&
\multicolumn{1}{|c|}{}\\
\multicolumn{1}{|c|}{}&
\multicolumn{1}{|c|}{}&
\multicolumn{1}{|c|}{$y_{0}$=0.3}&
\multicolumn{1}{|c|}{--}&
\multicolumn{1}{|c|}{-0.09 $\pm$ 0.02}\\
\hline
\end{tabular}
\newpage
Table 4. The number of experimental events ($N_{\rm{event}}$) and
$\pi^{-}$-mesons ($N_{\pi^{-}}$), the asyummetry coefficient ($\xi$),
the strength of the correlation ($\zeta$) and the mean rapidity
($<y>$) of $\pi^{-}$-mesons.\\
\   \par
\   \par
\   \par
\   \par
\   \par
\   \par
\begin{tabular}{|c|c|c|c|c|c|}    \hline
&    &     &    &     &   \\
$A_{P} - A_{T}$&$N_{\rm{event}}$ & $N_{\pi^{-}}$ & $\xi$ & $\zeta$ & $<y>$
\\
\hline
&    &     &    &     &   \\
CNe  & 723 &  6209$^{*}$  & -0.08 $\pm$ 0.02 & 0.85 $\pm$ 0.08
& 1.17
$\pm$ 0.06\\
\hline
&    &     &    &     &   \\
MgMg & 6239 & 50775  & -0.09 $\pm$ 0.02 & 0.84 $\pm$ 0.09 & 1.23
$\pm$ 0.07\\
\hline
&    &     &    &     &   \\
CCu  & 1866 &  12390  & 0.12 $\pm$ 0.03 & 1.29 $\pm$ 0.27 & 0.93
$\pm$ 0.06\\
\hline
&    &     &    &     &   \\
OPb  & 732 & 7023  & 0.23 $\pm$ 0.05 & 1.61 $\pm$ 0.35 & 0.73
$\pm$ 0.07\\
\hline
\hline
\end{tabular}
\par
\    \par
\     \par
\ $^{*}$ -- $\pi^{+}$ mesons are included also.
\newpage
\begin{figure} \begin{center}
\epsfig{file=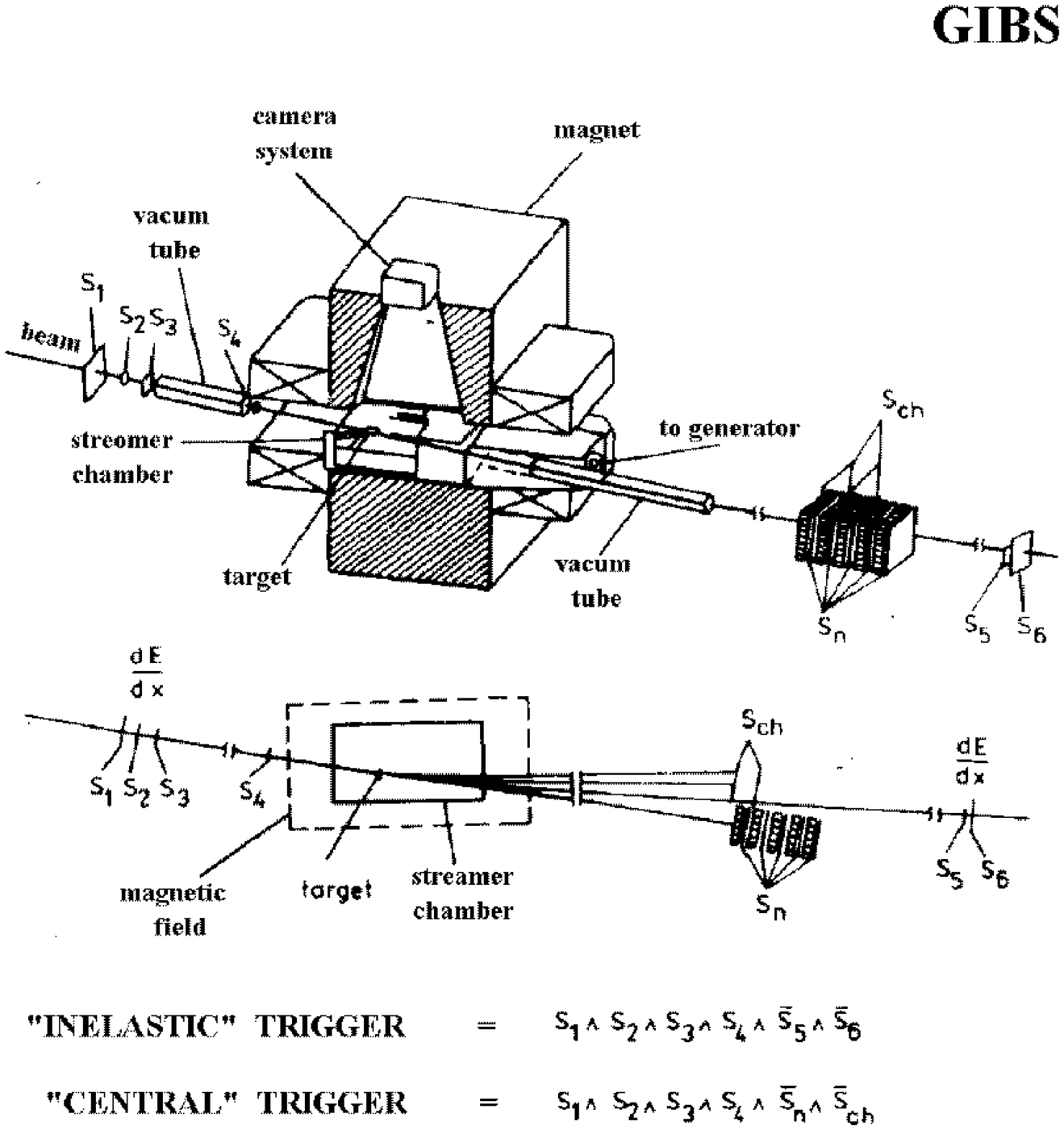,bbllx=0pt,bblly=0pt,bburx=594pt,bbury=842pt,
width=18.0cm,angle=0}
\end{center}
\vspace{-4.9cm}
\hspace{0.cm}
\begin{minipage}{16.0cm}
\caption
{ Experimental set-up. The trigger and trigger distances are
not to scale}
\end{minipage}
\end{figure}
\begin{figure}
\begin{center}
\epsfig{file=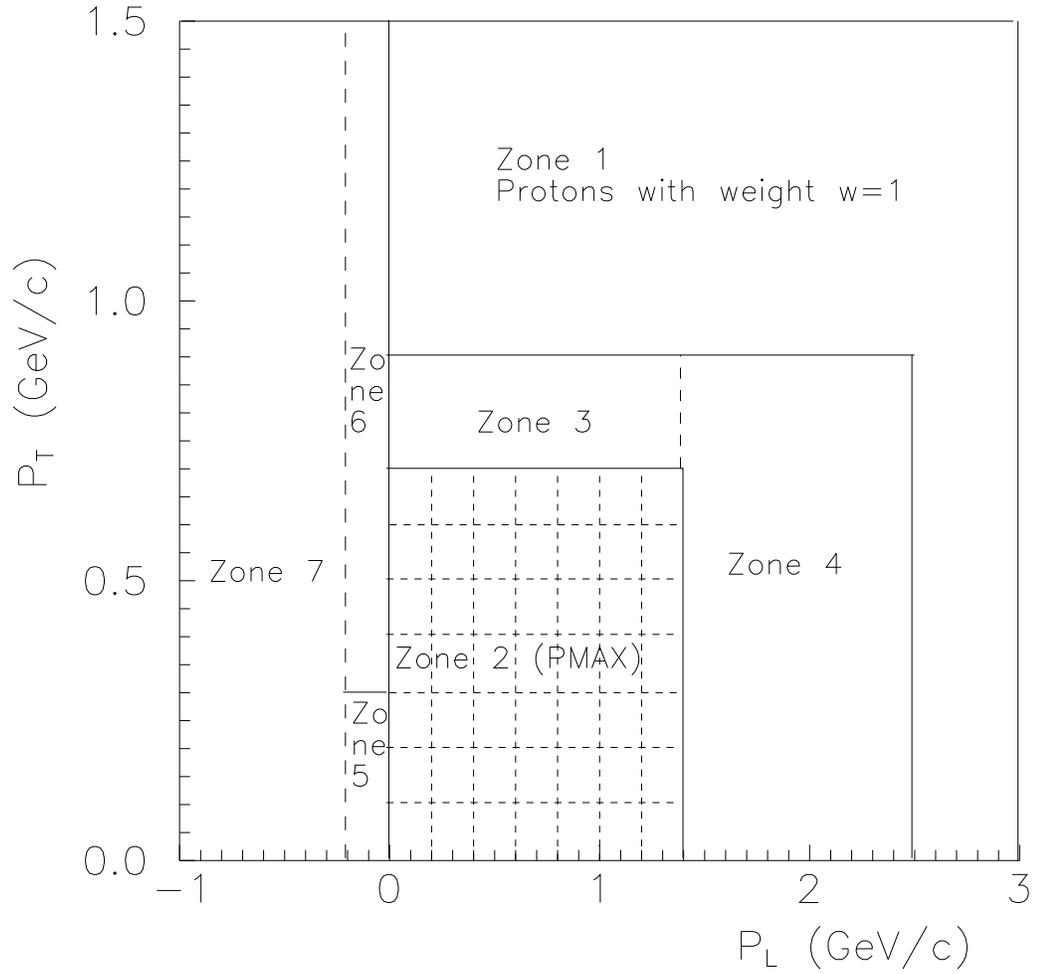,bbllx=0pt,bblly=0pt,bburx=594pt,bbury=842pt,
width=18cm,angle=0}
\end{center}
\vspace{-9.cm}
\hspace{0.cm}
\begin{minipage}{16.0cm}
\caption
{Two dimentional ($p_{T}$, $p_{L}$) ($p_{T}$ and $p_{L}$ denote
the transverse and longitudinal momenta, correspondently) plot
for the
in CNe collisions for the identifiction of protons and $\pi^{+}$ mesons.}
\end{minipage}
\end{figure}
\pagebreak
\begin{figure}
\begin{center}
\epsfig{file=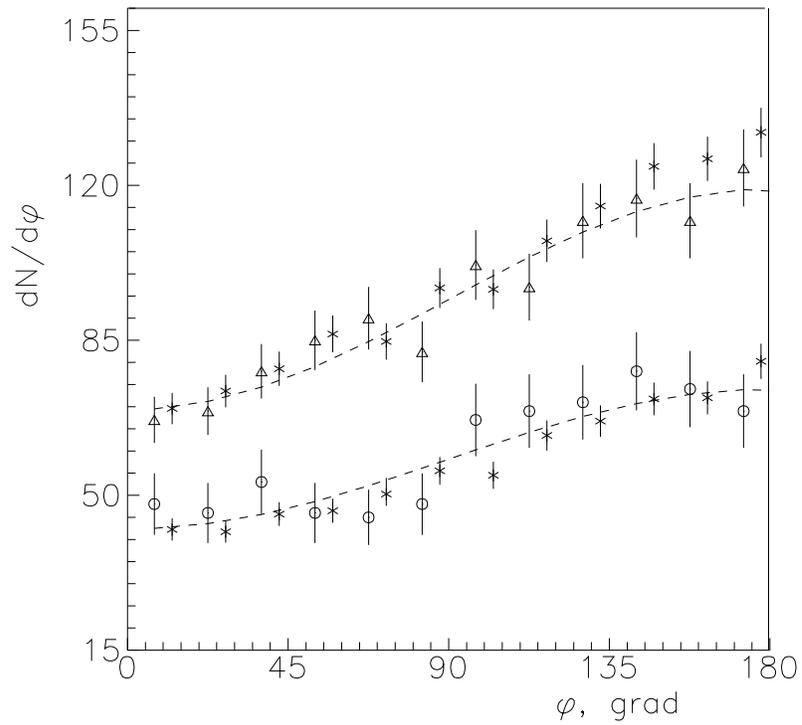,bbllx=0pt,bblly=0pt,bburx=594pt,bbury=842pt,
width=18cm,angle=0}
\end{center}
\vspace{-10.cm}
\hspace{3.cm}
\begin{minipage}{13.0cm}
\caption
{The dependence of the correlation function C($\Delta \varphi$) on the
$\varphi$
for protons from: ($\circ$) -- CNe and ($\bigtriangleup$) --  CCu
collisions.
({\scriptsize$\times$}) -- the QGSM data.
Dashed curves are the results of the approximation of the data (see
text).}
\end{minipage}
\end{figure}
\pagebreak
\begin{figure}
\begin{center}
\epsfig{file=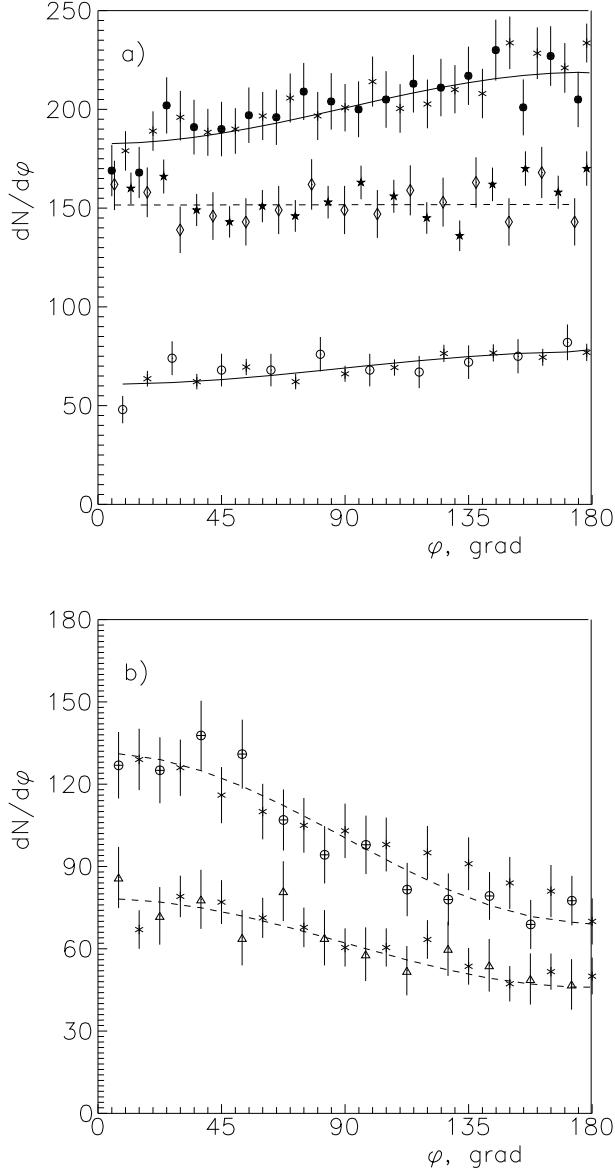,bbllx=0pt,bblly=0pt,bburx=594pt,bbury=842pt,
width=18cm,angle=0}
\end{center}
\vspace{-7.9cm}
\hspace{0.cm}
\begin{minipage}{16.0cm}
\caption
{The dependence of the correlation function C($\Delta$ $\varphi$) on the
$\varphi$
for $\pi^{-}$-mesons from:
(a) ($\circ$) -- CNe, ($\bullet$) -- MgMg ($n_{-}$ $\geq$ 7),
 ($\diamond$) -- MgMg ($n_{-}$ $<$ 7) and (b) ($\bigtriangleup$) -- CCu,
($\oplus$) -- OPb collisions.
 ({\scriptsize$\times$}, $\star$) -- the QGSM data, correspondently.
Solid and dashed curves are the results of the approximation of the
data
(see text).}
\end{minipage}
\end{figure}
\pagebreak
\begin{figure}
\begin{center}
\epsfig{file=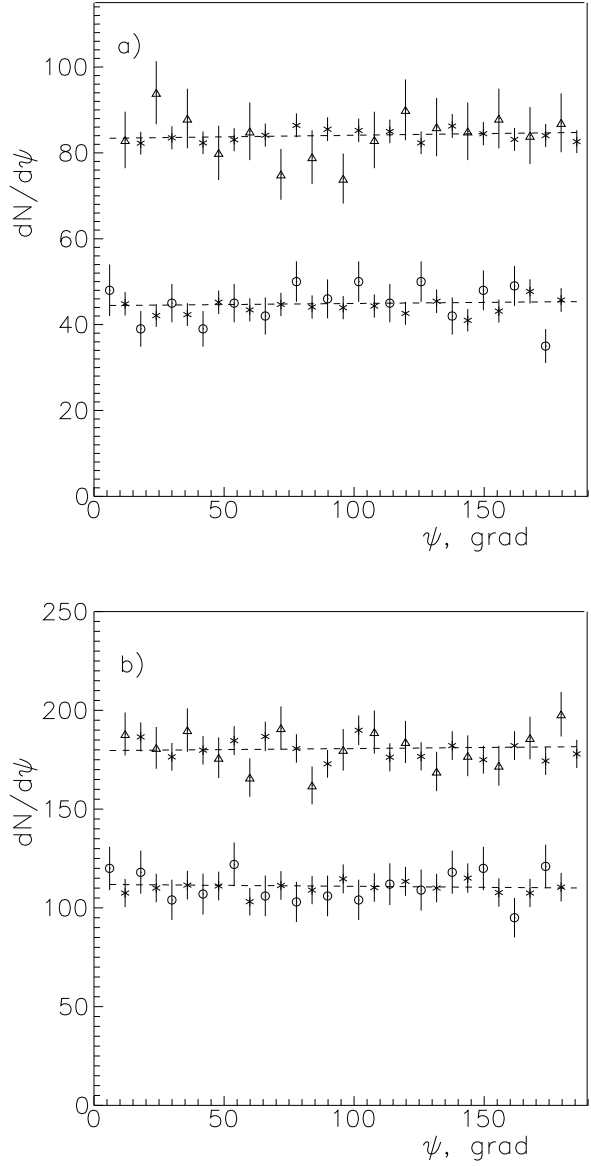,bbllx=0pt,bblly=0pt,bburx=594pt,bbury=842pt,
width=18cm,angle=0}
\end{center}
\vspace{-7.9cm}
\hspace{0.cm}
\begin{minipage}{16.0cm}
\caption
{The dependence of the correlation function C($\Delta \psi$) on the $\psi$
 in CCu collisions for protons (a) and for $\pi^{-}$-mesons (b).
 ($\circ$) -- for particles emitted in the backward hemisphere,
 ($\bigtriangleup$) -- for particles emitted in forward hemisphere.
 ({\scriptsize$\times$}) -- the QGSM data.
 Curves are the results of the approximation of the data by the first
order polynoms.}
\end{minipage}
\end{figure}
\pagebreak
\begin{figure}
\begin{center}
\epsfig{file=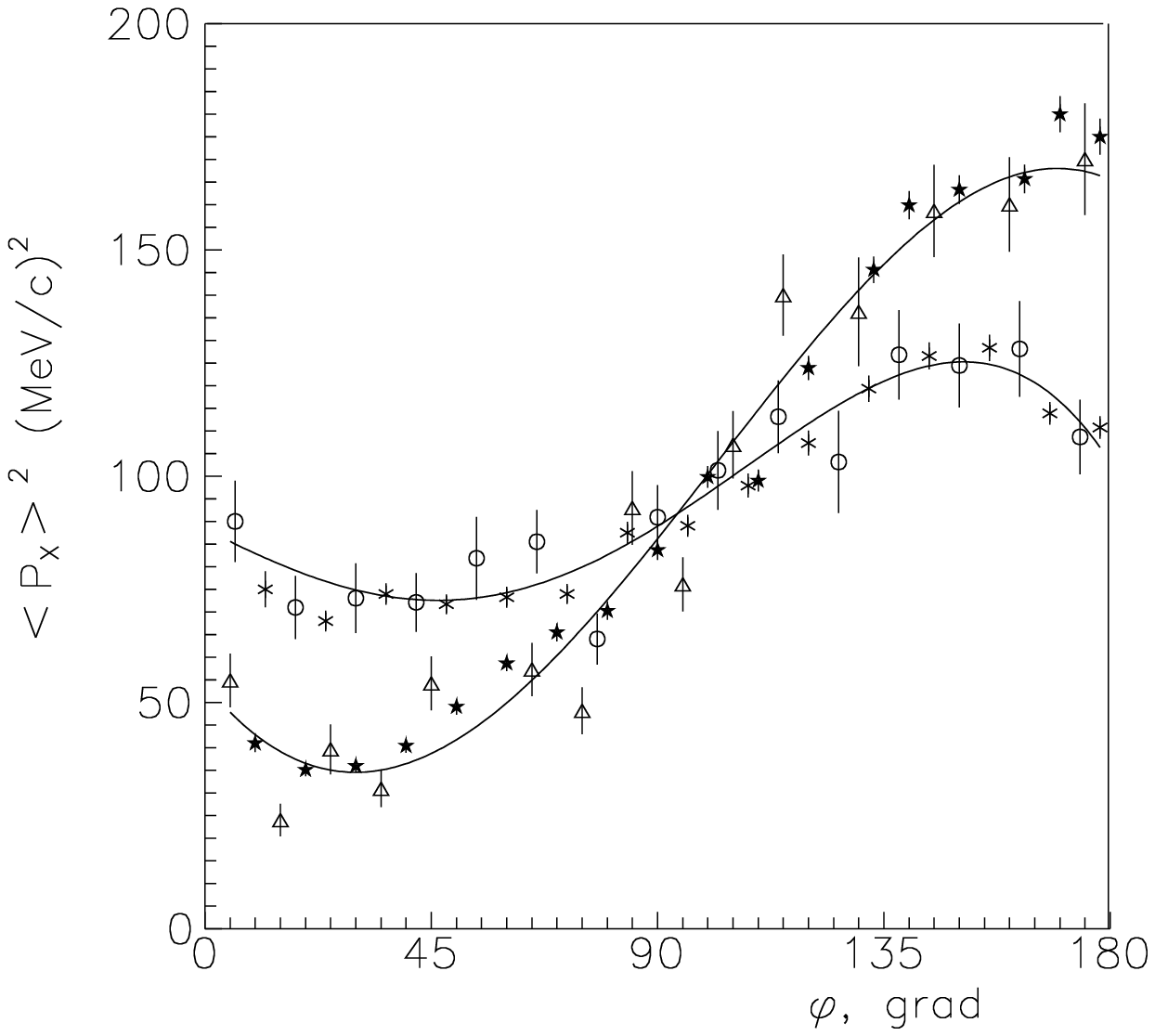,bbllx=0pt,bblly=0pt,bburx=594pt,bbury=842pt,
width=18cm,angle=0}
\end{center}
\vspace{-10.cm}
\hspace{3.cm}
\begin{minipage}{13.0cm}
\caption
{The dependence of $< P_{x} >^{2}$ on $\varphi$ (as described in the text)
for protons in: ($\circ$) -- CNe and ($\bigtriangleup$) -- CCu
collisions.
({\scriptsize$\times$}, $\star$) -- the QGSM data, correspondently.
Solid curves are the results of the approximation data by 4th
order polynoms.}
\end{minipage}
\end{figure}

\begin{thebibliography}{99}
\bibitem{}{M. Jacob and Van Than Tran, Phys. Rev. C {\bf{88}}, 725
(1980)}.
\bibitem{}{H. St$\ddot{o}$cker, {\it{et al}}., Phys. Lett. B {\bf{81}},
303
(1979)}.
\bibitem{}{H. A. Gustafsson, {\it{et al}}., Phys.Lett. B {\bf{142}}, 141
(1984)}.
\bibitem{}{M. Anikina, {\it{et al}}., Preprint No. E1-84-785, JINR (Dubna,
(1984)}.
\bibitem{}{M. Anikina, {\it{et al}}., Phys.Rev. C {\bf{33}}, 895 (1986)}.
\bibitem{}{S. Nagamiya and M. Gyalassy., Adv. in Nucl. Phys. {\bf{13}},
           201 (1984)}.
\bibitem{}{B. A. Li, W. Bauer, and G. Bertsch, Phys.Rev. C {\bf{44}}, 450
(1991)}.
\bibitem{}{M. D. Zubkov., Yad. Fiz. {\bf{55}}, 455 (1989)}.
\bibitem{}{N. Amelin, {\it{et al}}., Yad. Fiz. {\bf{52}}, 272 (1990)}.
\bibitem{}{N. Amelin, {\it{et al}}.,Phys. Rev. C {\bf{44}}, 1541 (1991)}.
\bibitem{}{V. V. Anisovich, {\it{et al}}., Nucl. Phys. B {\bf{133}}, 477
(1978)}.
\bibitem{}{H. R. Schmidt, {\it{et al}}., Nucl. Phys. A {\bf{544}}, 449
(1992)}.
\bibitem{}{T. C. Awes, {\it{et al}}., Phys. Lett. B {\bf{381}}, 29
(1996)}.
\bibitem{}{H. A. Gustafsson, {\it{et al}}., Z. Phys. A {\bf{321}}, 389
(1985)}.
\bibitem{}{P. Beckmann, {\it{et al}}., Modern Phys. Lett. A {\bf{2}}, 169
(1987)}.
\bibitem{}{B. P. Aduasevich, {\it{et al}}., Nucl. Phys. B {\bf{316}}, 419
(1990); \\
           B. P. Aduasevich, {\it{et al}}., Yad. Fiz. {\bf{57}} 268
(1994)}.
\bibitem{}{A. Kh. Vinitsky, {\it{et al}}., Yad. Fiz. {\bf{54}}, 12
(1991)}.
\bibitem{}{B. P. Aduasevich, {\it{et al}}., Yad. Fiz. {\bf{57}}, 2057
(1994)}.
\bibitem{}{H. Sorge, {\it{et al}}., Z. Phys. {\bf{644}}, 609 (1990)}.
\bibitem{}{Th. Lister, {\it{et al}}., Preprint No. 94-1, GSI (University
of Munster, 1994)}.
\bibitem{}{L. Chkhaidze, {\it{et al}}., Phys.Lett. B {\bf{411}}, 26
(1997)}.
\bibitem{}{L. Chkhaidze, {\it{et al}}., Phys.Lett. B {\bf{479}}, 21
(2000)}.
\bibitem{}{L. Chkhaidze, {\it{et al}}., nucl-ex/0008001, 2000; submitted
to Eur. Phys. J.}
\bibitem{}{L. Chkhaidze, {\it{et al}}., Eur. Phys. J. A {\bf{1}}, 2996
(1998)}.
\end{thebibliography}
\end{document}